\documentclass[aps,prx,twocolumn,showpacs,amsmath,amssymb,floatfix,superscriptaddress,notitlepage]{revtex4-1}
\usepackage{graphicx}
\usepackage{amsmath}
\newcommand{\ket}[1]{|{#1}\rangle}

\usepackage{amsfonts}
\usepackage{mathtools}

\usepackage[usenames,dvipsnames]{xcolor}
\usepackage[colorlinks=true,citecolor=Cerulean,linkcolor=RubineRed,urlcolor=Cerulean]{hyperref}

\begin{document}

\title{Synthetic mean-field interactions in photonic lattices}

\author{Callum W. Duncan}
\email{duncan@pks.mpg.de}
\affiliation{SUPA, Institute of Photonics and Quantum Sciences,
Heriot-Watt University, Edinburgh EH14 4AS, UK}
\affiliation{Max-Planck-Institut f\"{u}r Physik komplexer Systeme, D-01187 Dresden, Germany}
\author{Michael J. Hartmann}
\affiliation{SUPA, Institute of Photonics and Quantum Sciences,
Heriot-Watt University, Edinburgh EH14 4AS, UK}
\affiliation{Department of Physics, Friedrich-Alexander-Universit\"at Erlangen-N\"urnberg (FAU), 91058 Erlangen, Germany}
\author{Robert R. Thomson}
\affiliation{SUPA, Institute of Photonics and Quantum Sciences,
Heriot-Watt University, Edinburgh EH14 4AS, UK}
\author{Patrik \"Ohberg}
\affiliation{SUPA, Institute of Photonics and Quantum Sciences,
Heriot-Watt University, Edinburgh EH14 4AS, UK}

\begin{abstract}
Photonic lattices are usually considered to be limited by their lack of methods to include interactions. We address this issue by introducing mean-field interactions through optical components which are external to the photonic lattice. The proposed technique to realise mean-field interacting photonic lattices relies on a Suzuki-Trotter decomposition of the unitary evolution for the full Hamiltonian. The technique realises the dynamics in an analogous way to that of a step-wise numerical implementation of quantum dynamics, in the spirit of digital quantum simulation. It is a very versatile technique which  allows for the emulation of interactions that do not only depend on inter-particle separations or do not decay with particle separation. We detail the proposed experimental scheme and consider two examples of interacting phenomena, self-trapping and the decay of Bloch oscillations, that are observable with the proposed technique.
\end{abstract}

\maketitle

\section{Introduction}

Over recent years, photonic lattices have emerged as a fruitful platform for the investigation of states of matter \cite{christodoulides2003,Kitagawa2010,Garanovich2012,mukherjee2017,maczewsky2017,mukherjee2018,Mukherjee2018b}. Photonic lattices consist of periodic arrays of waveguides in either one or two dimensions, which are fabricated inside a glass substrate along its full length. The waveguides propagate in a single direction of the glass substrate which plays the role of time. The fabrication of a photonic lattice can be achieved using ultrafast laser inscription \cite{Davis1996}. The optical state in the photonic lattice behaves as if it was a single particle state in a tight-binding model, with a tunnelling and site-dependent potential. An advantage of photonic lattices is that the fabrication process allows for the dynamical control of the parameters of the tight-binding model. Many intriguing phenomena such as anomalous edge states in a periodically driven square lattice have been realised in photonic lattices by using the full control over the dynamical properties of the tunnelling \cite{mukherjee2017,maczewsky2017}. Photonic lattices have also been used to investigate quantum walks \cite{Peruzzo2010,Tang2018} and the photonic Zeno effect \cite{Biagioni2008}.

Photonic lattices have, so far, been restricted to the evolution of a state in relatively short timescales. This short timescale is normally fixed by the length of the glass substrate. However, recently state-recycling in a photonic lattice has been achieved \cite{mukherjee2018}. This technique allows for the output state to be reinserted phase coherently into the photonic lattice in a ring cavity scheme. The state-recycling allows for the physical phenomena associated with longer timescales to be realised and measured dynamically in photonic lattices. 

As photonic lattices deal with light they are usually characterised by their lack of interactions. In this work, we address this limitation by suggesting a technique to realise mean-field interactions. We propose to introduce synthetic interactions into the state-recycling set-up of Ref.~\cite{mukherjee2018} by emulating mean-field interactions to the optical state after each pass through the photonic lattice. The synthetic interaction is created by a conditional phase change which depends on the intensity of the light in the various lattice sites.

We will begin by revising the capability of photonic lattices to realise single-particle tight-binding models. We then motivate our proposed technique by comparing it to the Suzuki-Trotter decomposition utilised in numerics in Sec.~\ref{sec:Interactions}, followed by details of the proposed experimental set-up in Sec.~\ref{sec:Realisation}. We then consider two examples of realisable interaction dependent phenomena with the self-trapping of particles in the Josephson effect, and the decay of Bloch oscillations in a tilted lattice with a nonlinearity.

\section{Photonic lattices}

We envisage light propagating through an array of wave guides. Such a field is described by the discrete wave equation 
\begin{equation}
i\frac{dE_l}{dz}=-\kappa (E_{l+1}+E_{l-1})+\Delta\beta E_l \label{wg}
\end{equation}
where $E_{l}$ is the envelope of the electric field at the waveguide $l$, $z$ is the propagation direction,  $\kappa$ is the nearest-neighbour coupling constant, and $\Delta\beta$ the propagation constant. Eq. (\ref{wg}) is derived using the tight-binding and paraxial approximation, and is in the form of a Schr\"odinger equation if we identify the distance $z$ with the time $t$ and $\Delta\beta$ as a potential. This in turn allows us to also map the system to a Hubbard model 
\begin{align}
\hat{H}_0 = & - \sum_{i} J \left( \hat{b}^\dagger_{i+1} \hat{b}_{i} + \hat{b}^\dagger_{i} \hat{b}_{i+1}  \right) + \sum_{i} \epsilon_i \hat{n}_i,
\label{eq:GeneralHamiltonian}
\end{align}
with $\hat{n}_i$ the number operator, $\epsilon_i=\Delta\beta_i$ a  site dependent potential, and $J=\kappa$ the tunnelling strength.  The operators $\hat{b}^\dagger_{i}$ and $\hat{b}_{i}$ create and annihilate a particle at site $i$ respectively. However, with the mapping between the discrete wave equation and the Hubbard model it is important to realise that we are strictly emulating a single particle system where the light field takes the role of the wave function for the particle. We are consequently not emulating single photon dynamics; the beam contains after all many photons.  

The tunnelling strength $J$ can be modulated in the photonic lattice by controlling the distance between each waveguide. Periodically-driven models can be realised by varying the tunnelling or site-dependent potential along the length of the glass substrate. The effective time evolution of an optical input state to the photonic lattice will follow the unitary dynamics of
\begin{align}
\ket{\psi(t)} = e^{-i\hat{H}_0 t} \ket{\psi(0)},
\end{align}
where $\ket{\psi(0)}$ is the optical input state and $\ket{\psi(t)}$ the evolved state. In the photonic lattice, the single-particle tight-binding state can be written in the position representation as
\begin{align}
\ket{\psi(t)} = \sum_i \psi_i(t) \ket{i},
\end{align}
where $\psi_i(t)$ is the wave function at site $i$. 

The state recycling technique of Ref.~\cite{mukherjee2018} allows multiple passes of the optical state through a single photonic lattice by the use of standard optical components. This set-up, in effect, applies the unitary evolution under the general Hamiltonian in Eq. \eqref{eq:GeneralHamiltonian} multiple times. It allows for the observation of the state in between each unitary evolution using an array of single photon detectors where after each round trip a small fraction of the light beam is redirected to the detector array. With the detection of the light intensity in each lattice site, we then perform a conditional phase shift in each discrete lattice site, see Fig. 1. In Sec.~\ref{sec:Realisation}, we will extend this scheme to include mean-field interactions.

\section{Introducing interactions}
\label{sec:Interactions}

We are aiming to emulate two-body interactions which are of the general form 
\begin{align}
\hat{V} = \sum_{ij} U_{ij} \hat{n}_i \hat{n}_j,
\end{align}
with $U_{ij}$ denoting the interaction strength. For the Bose-Hubbard model the sum is often restricted to $i = j$, but, as we will show next, we are not restricted to this situation. The general tight-binding Hamiltonian is then $\hat{H} = \hat{H}_0 + \hat{V}$, with the unitary evolution over a time $t$ given by the operator
\begin{align}
e^{-i\hat{H} t} = e^{-i\left[\hat{H}_0 + \hat{V}\right] t}.
\end{align}
The photonic lattice itself implements the unitary time evolution with the original Hamiltonian $\hat{H}_0$ and we need to separate the evolution under $\hat{H}_0$ and $\hat{V}$. To first order the decomposition is
\begin{align}
e^{-i\left[\hat{H}_0 + \hat{V}\right] t} \approx \left( e^{-i \hat{V}t/n} e^{-i \hat{H}_0 t/n} \right)^n,
\end{align}
where the approximation is good for a large integer $n$. Therefore, we can implement the interacting dynamics by a separate unitary evolution, as long as the Suzuki-Trotter expansion is valid for which $\|\hat{H}_0 + \hat{V}\| \delta t\ll1$ with $\delta t=t/n$ small. In other words, we propose to realise the dynamics of the full Hamiltonian by applying a sequence of  a short evolution under only $\hat{H}_0$ and then a short evolution under only $\hat{V}$. This results in the full dynamics being realised in a step-wise fashion, similar to a numerical evolution of the state.

To realise interacting models in photonic lattices we propose that the photonic lattice itself enforces the evolution of the state under the unitary
\begin{align}
\hat{U}_{\mathrm{pl}} = e^{-i \hat{H}_0 \delta t},
\label{eq:UPhotonicLatt}
\end{align}
while by external optical means the unitary 
\begin{align}
\hat{U}_{\mathrm{ext}} = e^{-i \hat{V} \delta t},
\label{eq:UExternal}
\end{align}
is realised. These two unitary operations are then applied multiple times using the state-recycling technique \cite{mukherjee2018}.

While we evolve the optical state by a unitary operation of an interaction operator, the state remains in essence a single-particle state. We therefore interpret the process, i.e the conditional phase shift, as a mean-field effect. We take the number operator acting on the state to be defined by
\begin{align}
\hat{n}_j \ket{\psi(t)} =  |\psi_j(t)|^2 \ket{j}.
\end{align}
This means that the on-site two-body contact interactions for the Bose-Hubbard model for a site $j$ are proportional to $|\psi_j(t)|^4$ as expected \cite{lewenstein2012,pitaevskii2016}.

For this picture of mean-field interactions, the unitary interaction operator from Eq.~\eqref{eq:UExternal} is essentially a phase added to each individual lattice site where the phase shift in question depends on the light intensity in the lattice sites.  This would be implemented by measuring the light intensity after a single evolution through the photonic lattice and calculating the corresponding interaction term for each lattice site. With this information we can recreate the effect of $\hat{V}$ by imposing the required phase shifts. and allow implementation of $\hat{U}_{\mathrm{ext}}$. We will discuss the realisation of the interaction unitary operator further in the next section.

\section{Proposed realisation}
\label{sec:Realisation}

\begin{figure}[t!]
\begin{center}
\includegraphics[width=0.5\textwidth]{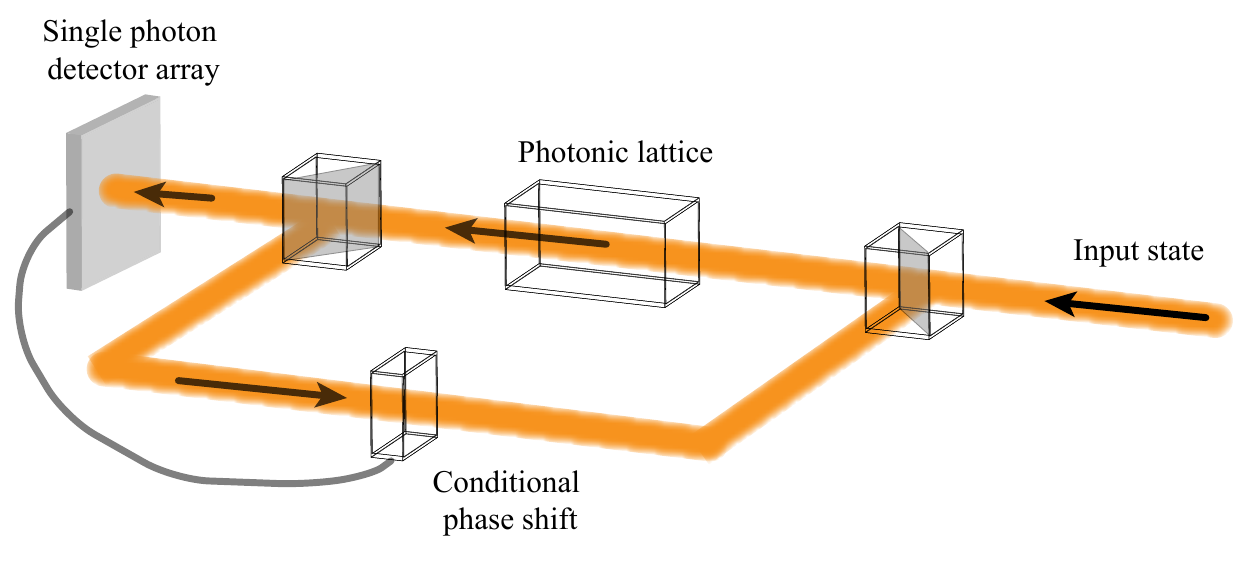}
\end{center}
\caption{Simplified illustration of the proposed experimental technique of mean-field interacting photonic lattices, with the photonic lattice implementing the unitary evolution of Eq.~\eqref{eq:UPhotonicLatt} together with a conditional phase shift to implement the unitary operator in Eq.~\eqref{eq:UExternal}.}
\label{fig:SetUpIllustration}
\end{figure}

A simplified illustration of our proposed experimental technique is shown in Fig.~\ref{fig:SetUpIllustration}. This set-up builds on what has already been realised experimentally for state-recycling in photonic lattices \cite{mukherjee2018}. We propose to add to the established state-recycling experimental technique an extra stage for the optical state to be modified. This extra stage, which provides a conditional phase shift to the light, is required to apply the unitary operator representing the interactions given by Eq.~\eqref{eq:UExternal}. The interaction operator adds a phase for each lattice site which is dependent on the intensity of the light in the lattice sites. One possible method to implement such a unitary operator and phase shifts would be to use electro-optic phase modulators which can address each lattice site. 

An advantage of the state-recycling technique is that it allows the dynamical imaging of the optical state. This is achieved by passing part of the optical intensity to an array of single photon detectors whose numbers match the number of lattice sites, as illustrated in Fig.~\ref{fig:SetUpIllustration}. In order to implement the mean-field interaction operator of Eq.~\eqref{eq:UExternal}, knowledge of the occupation, i.e. intensity of the light, at each lattice site is required. Therefore, we envisage a dynamical feedback from the imaging of the intensity to the device which in turn creates the required phase shift which emulates the interaction operator. In the numerics of Sec.~\ref{sec:Josephson} and \ref{sec:Bloch} we update the interaction unitary on every pass through the state-recycling system. 

The state-recycling technique reported in \cite{mukherjee2018} clearly demonstrated the basic concept of state-recycling and time-resolved imaging, but it should also be acknowledged that there are a number of challenges that must be overcome to experimentally implement the full theoretical framework outlined in this paper. The first of these is optical loss. Ultrafast laser inscribed waveguides are low index contrast waveguides, meaning that photonic lattices containing multiple waveguide bends are generally long (e.g. $\sim$ $10$ cm). Furthermore, even state-of-the-art ultrafast laser inscribed waveguides exhibit propagation losses of $\sim 0.5$ dB/cm at $800$ nm. These two factors mean that signal loss will rapidly accumulate and prevent useful measurements after only a relatively low number of round trips unless steps are taken. To address this, we envisage state-recycling experiments employing loss compensating gain provided by an optical amplifier. This amplifier could be based on a multicore optical fibre with an array of cladding-pumped rare-earth doped guiding cores to simultaneously amplify many independent spatial modes \cite{Suzuki:07}. The second experimental challenge that must be overcome is to devise a way to precisely modulate the relative phase of each mode in the photonic lattice in an arbitrary manner for each round trip. This is particularly challenging if the round-trip time of the state-recycling cavity is only a few ns, as was the case in \cite{mukherjee2018}. Spatial light modulators based on liquid crystals provide unparalleled phase control of thousands of independent pixels but can only deliver refresh rates in the few kHz range. Therefore, they are only suitable for controlling the round-trip phase properties of the static photonic lattice cavity, but not for implementing the mode intensity conditional phase modulation after each round trip. Another platform that is currently available is MEMS-based phase modulators \cite{tzang2019} which can achieve refresh rates of $\sim$ 0.5 MHz, and as such could be suitable for photonic lattice cavities with round trip times on the order of $\sim$ $\mu$s. Such long round-trip times are not impossible using free space optics \cite{Papadopoulos:04} and could be relatively easy to implement using fibre delay lines based on polarization maintaining multicore fibres \cite{Stone:14}. Finally, it should also be highlighted that new phase-modulator technologies are being actively pursued \cite{peng2019}, which could provide GHz refresh rates.


Our proposed technique allows for the control of the ratio of system parameters without the need to swap out optical components. The changing of optical components can result in increased experimental set-up time, due to the aligning of all components. For interacting systems it is usual to consider as a parameter the ratio of the tunnelling and interaction strengths. The proposed set-up allows for simple access to this parameter by only modulating the strength of the phase term applied by the external optics.

\section{Examples of realisable interacting phenomena}

\subsection{Josephson effect}
\label{sec:Josephson}

A well known phenomenon in interacting systems is the Josephson effect \cite{Josephson1962} which has been observed in superconductors \cite{Likharev1979}, superfluid helium \cite{pereverzev1997}, and ultracold gases \cite{Albiez2005}. The Josephson effect is essentially a collective coherent tunnelling of multiple particles through a large barrier which is driven by a quantum phase difference between the state either side of the barrier. For two BECs trapped in a double well the dynamics are well described by the Gross-Pitaeveskii equation \cite{Smerzi1997,Raghavan1999} and oscillations in the imbalance of the occupation of the two wells are observed. These oscillations correspond to the usual ac current produced from a dc voltage in superconducting Josephson junctions \cite{Smerzi1997,pitaevskii2016,larkin2009}. In the case of ultracold gases the dc voltage is analogous to the initial population imbalance of the double well potential. 

For the Josephson effect with a nonlinearity due to interactions there are two clear and distinct regimes \cite{Smerzi1997,Albiez2005,bloch2005,Salgueiro2007,abbarchi2013}. For small initial imbalance of the occupation of the lattice sites or small interaction strength, there are oscillations in the imbalance corresponding to the alternating flow of atoms between the sites. However, for large initial imbalance or strong interactions, the system exhibits a self-trapping in which the population imbalance no longer oscillates with a mean of zero imbalance. 

\begin{figure}[t]
\begin{center}
\includegraphics[width=0.45\textwidth]{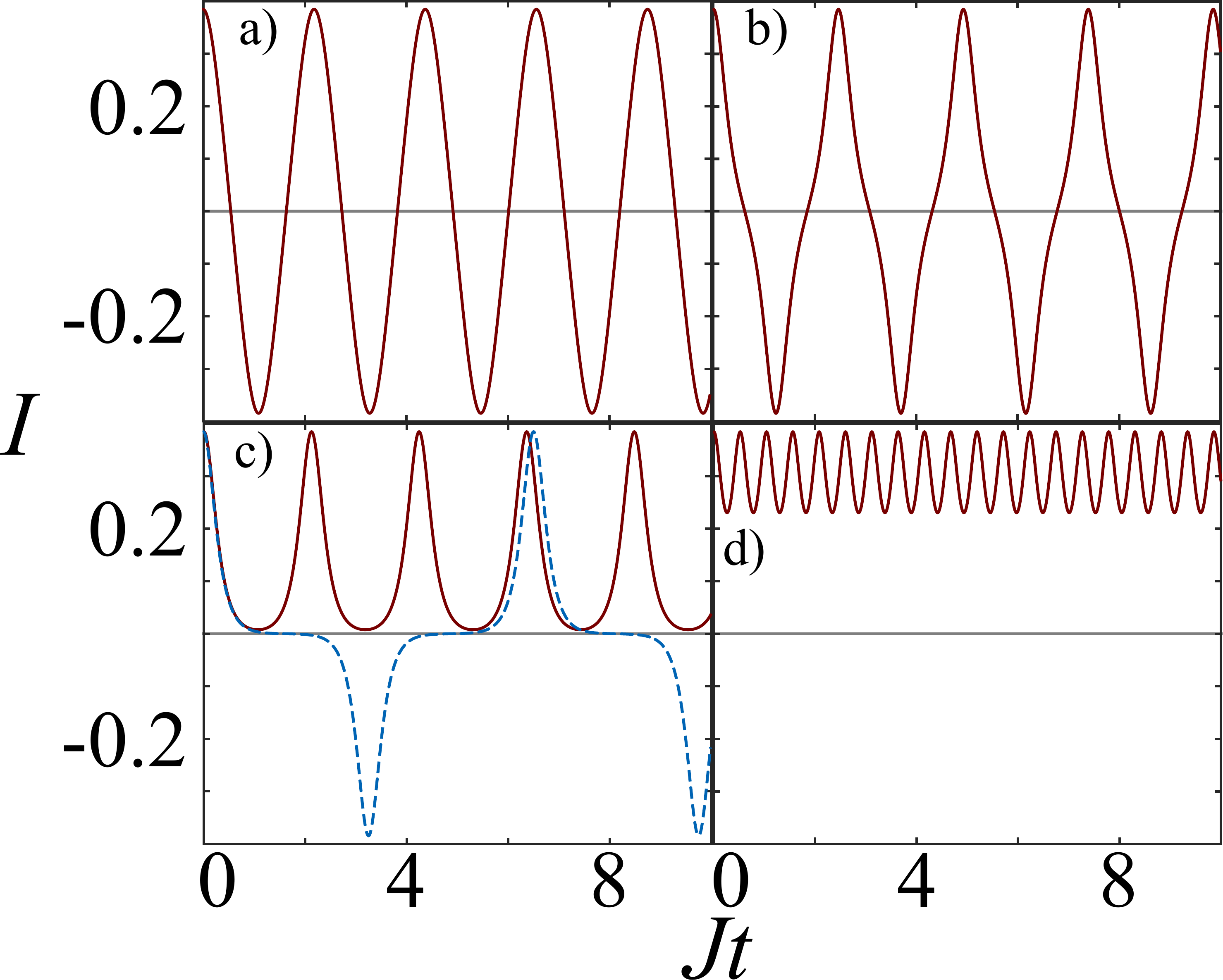}
\end{center}
\caption{Population imbalance $I(t)$ showing the oscillating flow of population between sites. Initial condition of $I(0) = 0.38$ with $\delta t = 0.01$, $10^3$ state-recycling steps and interaction strengths of a) $U/J=10$, b) $U/J = 25$, c) $U/J=26$ (dashed line $U/J = 25.99$), and $U/J = 40$. }
\label{fig:JosephOscillations}
\end{figure}

\begin{figure}[t]
\begin{center}
\includegraphics[width=0.48\textwidth]{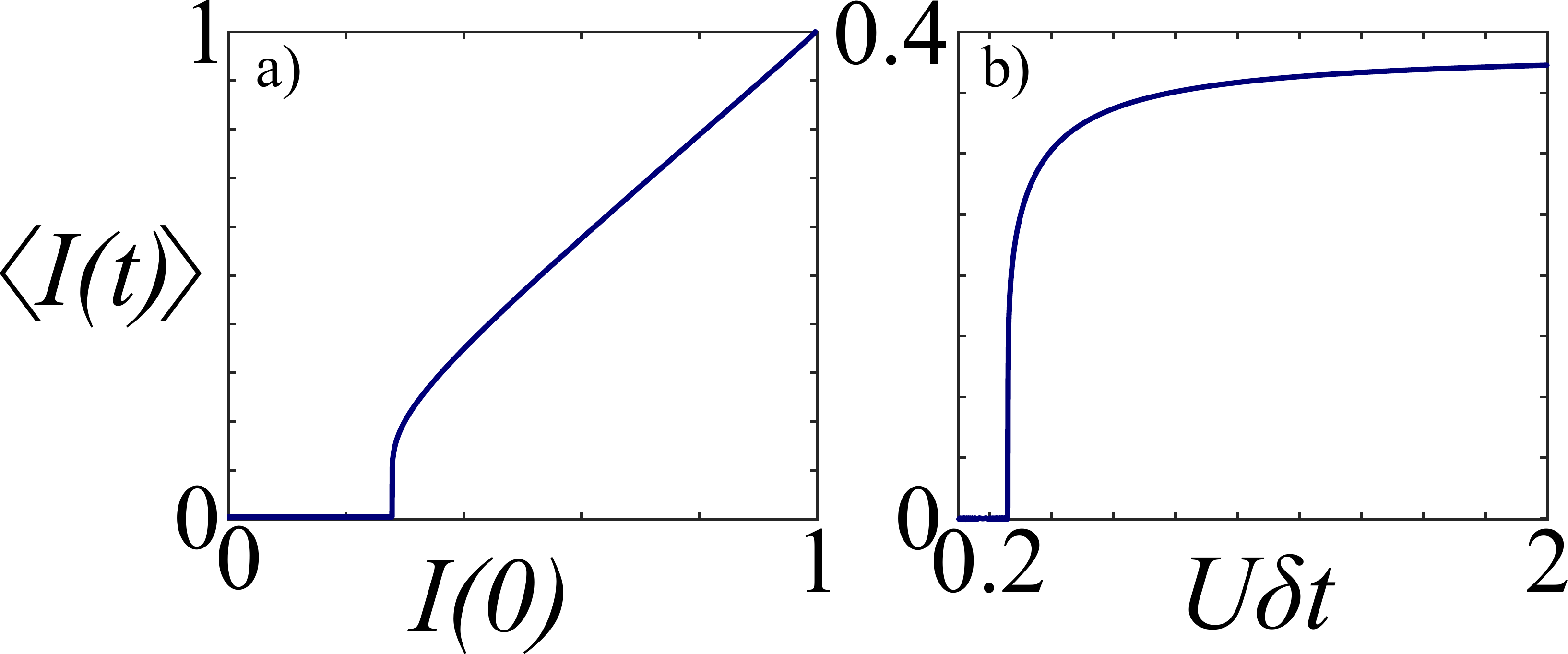}
\end{center}
\caption{Average population imbalance $\langle I(t) \rangle$ showing the sharp transition between the oscillating and self-trapping regimes. a) Across initial imbalance $I(0)$ with $U/J = 50$, and b) across $U\delta t$ with $I(0) = 0.38$. Time-step of $\delta t = 10^{-3}$, and the average is across $2\times10^{4}$ state-recycling steps.}
\label{fig:MeanImbalanceTransition}
\end{figure}

The double-well Josephson junction can be described by the Hamiltonian
\begin{align}
\hat{H} = -\frac{J}{2} \left( \hat{b}^\dagger_{1} \hat{b}_2 +  \hat{b}^\dagger_{2} \hat{b}_1 \right) + U \left( \hat{n}_1 \hat{n}_1 + \hat{n}_2 \hat{n}_2 \right),
\label{eq:JosephsonH}
\end{align}
with the sites of the double well being labelled as $1$ and $2$, $J$ being a positive tunnelling strength, and $U$ a positive interaction strength (i.e. a repulsive interaction). We will work in units of energy in terms of $J$ and time in terms of $J^{-1}$. The population imbalance in the mean-field picture is defined as
\begin{align}
I = |\psi_1|^2 - |\psi_2|^2,
\end{align}
with $\psi_{1(2)}$ being the coefficients of the state in site $1(2)$.

As in the BEC experiment \cite{Albiez2005}, we initialise the optical state with an initial imbalance, then we observe the dynamics of the system by the unitary evolution of the state under Hamiltonian~\eqref{eq:JosephsonH}. The numerics are achieved with the Suzuki-Trotter decomposition approach discussed in Sec.~\ref{sec:Interactions}, where we apply the unitary evolution of the photonic lattice and interaction seperately. We evolve for $10^3$ state-recylcing steps with a time step $\delta t = 10^{-3}$. Increasing $\delta t$ to $10^{-2}$ and considering $10^2$ state-recycling steps results in the same behaviour being observed but with less detail. We observe the expected Josephson oscillations for weak interactions in Figs.~\ref{fig:JosephOscillations}a and b. There is then the expected transition point for intermediate interactions between the two regimes shown in Fig.~\ref{fig:JosephOscillations}c. The self-trapping regime is clearly observed for large interaction strengths and is shown in Fig.~\ref{fig:JosephOscillations}d. This self-trapping is counterintuitive as a repulsive system is expected to delocalise and occupy both sites equally to minimise its energy. This is a clear sign of the presence of interactions in the system. Note the striking similarity between Fig.~\ref{fig:JosephOscillations} and the dynamics of two BECs described by two coupled Gross-Pitaevskii equations in Ref.~\cite{Smerzi1997}. 

\begin{figure}[t!]
\begin{center}
\includegraphics[width=0.48\textwidth]{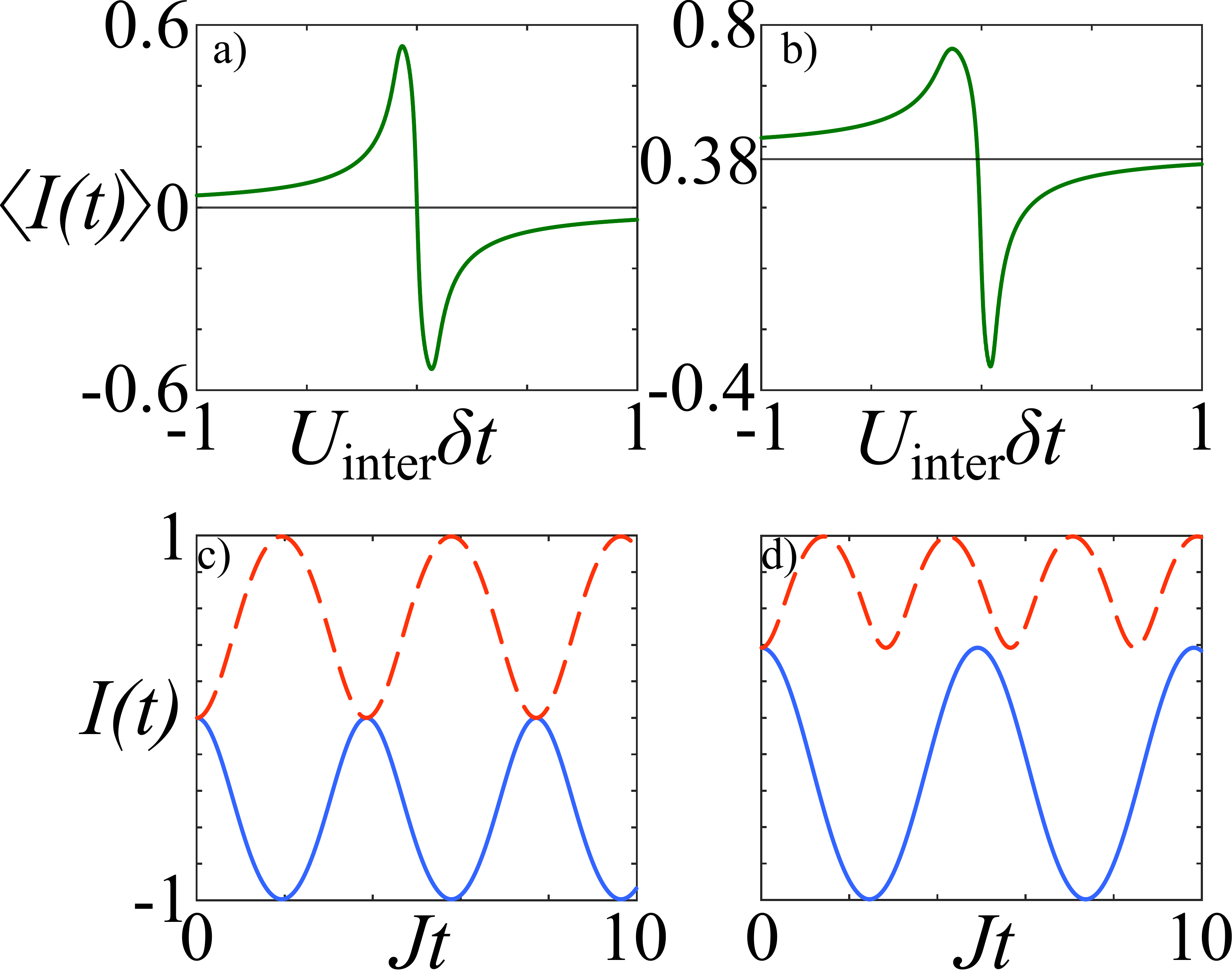}
\end{center}
\caption{Average population imbalance $\langle I(t) \rangle$ with inter-site interaction showing a transition between oscillating and self-trapping regimes. The on-site interaction is set to zero, the time-step of $\delta t = 10^{-3}$, and the average is across $2\times10^{4}$ state-recycling steps. a) Initial imbalance of $I(0)=0$ and b) initial imbalance of $I(0) = 0.38$. c) The self-trapping oscillations for the maximum (minimum) $\langle I(t) \rangle$ at $U_{\mathrm{inter}}\delta t = -0.066 (0.066)$ is shown by a dashed red (solid blue) line for $I(0)=0$. d) The self-trapping oscillations for the maximum (minimum) $\langle I(t) \rangle$ at $U_{\mathrm{inter}}\delta t = -0.135 (0.0039)$ is shown by a dashed red (solid blue) line for $I(0)=0.38$.}
\label{fig:UInterMean}
\end{figure}

Photonic lattices lend themselves well to experiments requiring many repetitions, hence we can consider measuring the average imbalance for a range of interaction strengths or initial imbalances to fully characterise this transition as is shown in Fig.~\ref{fig:MeanImbalanceTransition}. We take the average imbalance across $2\times10^4$ state-recycling steps. Note, that for smaller numbers of state-recycling steps, e.g. $10^2$, the characteristics of the transition are still captured but with small oscillations around the plateaus shown in Fig.~\ref{fig:MeanImbalanceTransition}. For the case of increasing the interactions to go across the transition, the average imbalance asymptotically tends to the value of the initial imbalance for large interactions. 

As the interactions are implemented by optical means, we are not limited to considering only on-site interactions, and we are able to consider the interactions to be highly position dependent. We are also not constrained by the physical conditions of interactions in cold atoms or superconducting systems \cite{jin2013,collodo2019}. Therefore, we can consider the asymmetric case of the Josephson effect by introducing a term of the form
\begin{align}
\hat{V}_{\mathrm{inter}} = U_{\mathrm{inter}} \hat{n}_1 \hat{n}_2 \delta_{i,1},
\label{eq:NonSymmInter}
\end{align}
where $U_{\mathrm{inter}}$ is the strength of this inter-site interaction and $\delta_{i,1}$ is the Kronecker delta. This interaction means that the population of the first site of the double well `feel' the presence of the population of the second site. Such an interaction term would be unphysical in the framework of ultracold atoms, as this would be a contact interaction term of one condensate with the other without the reciprocal interaction. This consideration of only one of the populations `feeling' an interaction is analogous to having a system where the respective intersite couplings have a large energy scale difference. This would allow for one of the interaction terms to be removed in a similar vein to the Born-Oppenheimer approximation \cite{born1924}.

Taking the Josephson Hamiltonian~\eqref{eq:JosephsonH} and adding the non-symmetric interaction term of Eq.~\eqref{eq:NonSymmInter} results in some interesting behaviour. We consider the case of the usual interactions being turned off, i.e. $U=0$, and initial imbalances of $I(0) = 0$ and $I(0)=0.38$. We then sweep across attractive and repulsive strengths of the non-symmetric inter-site interaction. We observe that for small $U_{\mathrm{inter}} \delta t$, the system becomes self trapped, with a fast movement away from $\langle I(t)\rangle = 0$. The population is trapped in either site $1$ or $2$ depending on if the interaction is attractive or repulsive. For intermediate interactions there is a peak in the absolute average imbalance. The maximum and minimum average imbalances occur when the population is oscillating between its initial value and the full population being in a single site, as is shown in Figs.~\ref{fig:UInterMean}c and d. With this defining the extreme points of the average imbalance it is of no surprise that the case of $I(0)=0$ is symmetric and $I(0)\neq 0$ is not. For large inter-site interactions, $\langle I(t)\rangle$ tends towards the initial imbalance. Which for the case of $I(0)\neq 0$ means for large interaction strengths the population is self trapped much like before which is again counterintuitive to simple energy arguments.

\subsection{Bloch oscillations}
\label{sec:Bloch}

It is well known that a single particle confined on a lattice and experiencing a constant force will exhibit Bloch oscillations \cite{kittel1976}. In ultracold atoms in optical lattices Bloch oscillations have been observed \cite{Dahan1996,Anderson1998}, including recently in position space by using absorption imaging \cite{Geiger2018}. Bloch oscillations have also been observed in photonic lattices \cite{mukherjee2018}, including for the case of two interacting particles in one dimension by extending the model to a non-interacting two-dimensional problem \cite{corrielli2013}. Bloch oscillations are a good observable for characterising the superfluid to Mott-insulator transition, as they are overdamped for the Mott-insulator phase \cite{Gorshkov2011,Carrasquilla2013}. In general, for the interacting Bose-Hubbard model, it is known that Bloch oscillations decay because of the introduced nonlinear dephasing and revivals of the oscillations at later times can occur \cite{Witthaut2005,Kolovsky2009,Mahmud2014}.

\begin{figure}[t!]
\begin{center}
\includegraphics[width=0.48\textwidth]{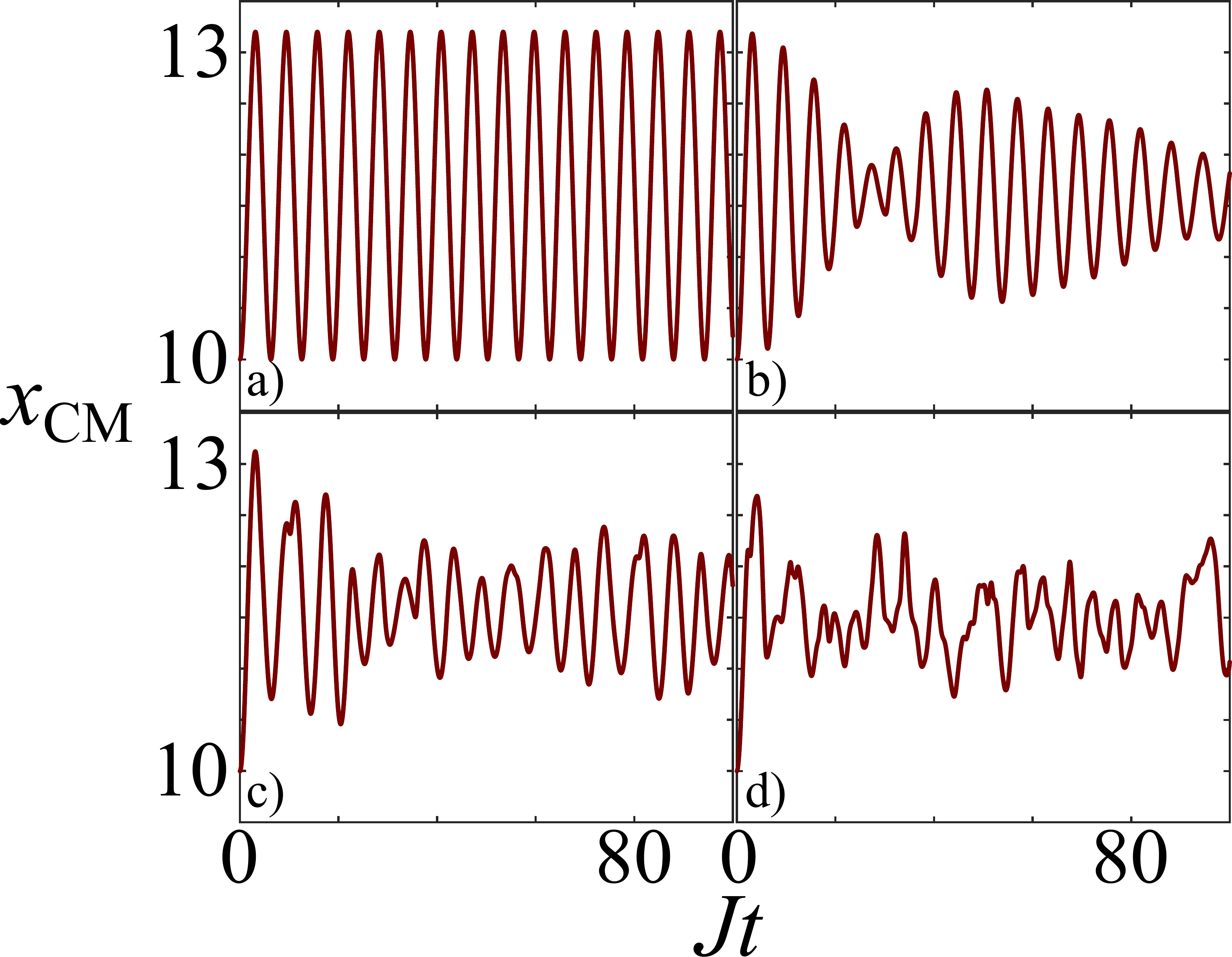}
\end{center}
\caption{Centre-of-mass motion $x_{\mathrm{CM}}$ exhibiting Bloch oscillations for a lattice of size $L=21$, with initial equal occupancy of the middle five sites, time step $\delta t = 0.01$, and $2 \times 10^4$ total round trips of evolution. a) non-interacting case, b) $U/J = 5$, c) $U/J	= 10$, and d) $U/J = 50$.}
\label{fig:BOU}
\end{figure}

We will consider a one-dimensional Bose-Hubbard model with the addition of a force that tilts the lattice. The Hamiltonian enforced by the photonic lattice is then of the form
\begin{align}
\hat{H}_0 = -J \sum_i \left( \hat{b}^\dagger_{i+1} \hat{b}_{i} + \hat{b}^\dagger_{i} \hat{b}_{i+1}  \right) - \Omega \sum_{i} x_i \hat{n}_i,
\end{align}
with $J$ a constant tunnelling strength, $\Omega$ the strength of the tilt, and $x_i$ the label of the $i$th lattice site. The conditional phase shift then realises the mean-field on-site contact interaction
\begin{align}
\hat{V} = \frac{U}{2} \sum_{i} \hat{n}_i \hat{n}_i,
\end{align}
with $U$ being the interaction strength. We will again work in units of the tunnelling strength $J$, and we will set $\Omega = 1$. We will study the centre-of-mass motion 
\begin{align}
x_{\mathrm{CM}} (t) = \frac{1}{N} \sum_i^L x_i \langle \hat{n}_i \rangle
\end{align}
where $N$ is the normalisation of our state, and the sum is over the full lattice size $L$.

We consider an example of a $21$ site lattice with an initial state of equal occupancy of the middle five sites. We have avoided any occupancy near the edge of the system to remove any finite-size effects. The state is evolved for a total of $2 \times 10^4$ state-recycling steps with a time step of $\delta t = 10^{-3}$. We plot the centre-of-mass motion in Fig.~\ref{fig:BOU}, where we consider the non-interacting, weakly interacting, and strongly interacting regimes of the system. For the case of no interactions, see Fig.~\ref{fig:BOU}a, we observe the usual robust Bloch oscillations. For weak interactions, see Fig.~\ref{fig:BOU}b, we observe the expected decay and revival of the Bloch oscillations.  It is known that for strong enough interactions the Bloch oscillations start to exhibit classical chaos \cite{Buchleitner2003,Thommen2003}, and we do observe the transition to the chaotic regime in Figs.~\ref{fig:BOU}c and d. The chaos in the oscillations come about due to the nonlinearity introducing occupation dependent frequencies to the motion which blur out the Bloch oscillations.

\section{Conclusions}

We have proposed a technique to introduce synthetic mean-field interactions in photonic lattices. The technique implements the interaction term independently from the photonic lattice and is inspired by the Suzuki-Trotter decomposition. By measuring the light intensity at each round trip it is possible to implement a mean-field interaction by a conditional phase shift for each lattice site. The proposed technique allows for mean-field interactions of arbitrary range and strength to be implemented. The simulation of the full Hamiltonian is analogous to a step-wise implementation of the dynamics. We have shown that for two examples, the Josephson effect and Bloch oscillations, the expected phenomena of interactions could be observed in photonic lattices with the proposed technique of this work. The proposed inclusion of mean-field interactions in photonic lattices paves the way towards further development and simulation of interacting models in this robust platform.


\acknowledgements{
C.W.D. acknowledges support from EPSRC CM-CDT Grant No. EP/L015110/1. M.J.H. acknowledges support by Google Research, where his contribution to the work was finalized during a visiting faculty appointment.}



%

\end{document}